
%
%
\headline={\ifnum\pageno=1\firstheadline\else
\ifodd\pageno\rightheadline \else\leftheadline\fi\fi}
\def\firstheadline{\hfil}
\def\rightheadline{\hfil}
\def\leftheadline{\hfil}
\footline={\ifnum\pageno=1\firstfootline\else
\otherfootline\fi}
\def\firstfootline{\rm\hss\folio\hss}
\def\otherfootline{\hfil}
\def\lb{\hfil\break}
\def\to{\rightarrow}
\def\del{\partial}

\def\sqr#1#2{{\vcenter{\hrule height.#2pt\hbox
       {\vrule width.#2pt height#1pt\kern#1pt
       \vrule width.#2pt}\hrule height.#2pt}}}

\def\ddp{{d^dp \over {(2\pi)^d}}}
\def\ddk{{d^dk \over {(2\pi)^d}}}
\def\phivec{\vec\phi}
\def\Phivec{\vec\Phi}

\def\ambj{{1}}
\def\zinn{{3}}\def\brez{{4}}
\def\neub{{5}}\def\brezA{{6}}\def\schn{{7}}
\def\bm{{8}}\def\divec{{9}}
\def\mosh{{10}}\def\thooft{{11}}\def\eyal{{12}}
\font\twelvebf=cmbx10 scaled\magstep 1
\font\twelverm=cmr10 scaled\magstep 1
 1

\font\tenbf=cmbx10
\font\tenrm=cmr10
\font\tenit=cmti10

\parindent=1.5pc
\hsize=6.0truein
\vsize=8.5truein
\nopagenumbers
 \centerline{\tenbf   O(N) VECTOR MODELS
IN THE LIMIT $g \to g_c$ AND FINITE TEMPERATURE$^{*}$}
\vglue 0.8cm
\centerline{\tenrm MOSHE MOSHE $^{\dagger}$}
\baselineskip=13pt
\centerline{\tenit Physics Department, Technion - Israel
Inst. of Technology}
\baselineskip=12pt
\centerline{\tenit Haifa, 32000  ISRAEL}
\baselineskip=13pt
\vglue 0.8cm
\centerline{\tenrm ABSTRACT}
\vglue 0.3cm
{\rightskip=3pc
 \leftskip=3pc
 \tenrm\baselineskip=12pt\noindent
In the limit where
$N\to\infty$ and the coupling constant
$g \to g_{c}$ in a correlated
manner,  O(N) symmetric vector models  represent
filamentary surfaces.
The purpose of these studies is to gain
intuition for the long lasting search
for a possible description of quantum field theory
in terms of extended objects.
It is shown here that a certain limiting
procedure has to be followed in order to
avoid several difficulties
in establishing the theory at a critical
negative coupling constant.
It is also argued that at finite temperature a certain
metastable-false vacuum disappears as the
temperature is increased.
\vglue 0.6cm}
\vfil
\twelverm
\baselineskip=14pt
\vglue 1pt
{}From a phenomenological viewpoint it is
desirable to be able to describe short and
long distance  hadronic phenomena
by very distinct descriptions of quantum field
theory:(1) local quanta
(2) extended objects, respectively. The  understanding
of quantum field theories in terms of
extended objects is, however, a long lasting
problem in elementary
particles theory.  In the limit, where
$N\to\infty$ and the coupling constant
$g \to g_{c}$ in a correlated
manner,  O(N) symmetric vector models  represent
filamentary surfaces
$^{\ambj-\zinn}$ ~-~ randomly branched polymers. This is
in the same manner as matrix models
provide representations  of
dynamically triangulated random
surfaces in their double scaling limit $^{\brez}$ .
The surfaces and polymer chains are represented by
the  Feynman graphs  of the matrix and
vector  model, respectively .
 The double scaling limit in O(N) vector quantum field theories
reveals  an interesting new phase structure, as was
argued also in the case of matrix models.
 Though  matrix theories attract most attention,
a detailed understanding
of the extended objects description of these theories exists, to
some extent, only for dimensions  $d\leq 2$. On the other hand,
in  many cases, the O(N)
vector models can be successfully studied$^{\zinn}$ also  in
dimensions $d\geq 2$ and thus gaining intuition for the search
for a possible description of quantum field theory
in terms of extended objects in four dimensions
(see for example Ref.${\neub}$).

At zero temperature,  the self
interacting scalar O(N)
symmetric vector model in d  Euclidian dimensions
 is defined by the functional integral
$$ Z_N=
\int {\cal D}\Phivec \, exp \bigl\{-
\int d^d x \,\{~ {1\over2}{(\partial_{\mu} \Phivec)}^2 +
{\mu^2_0\over 2}\Phivec^2  +
{\lambda_0 \over 4}(\Phivec^2)^2\}~~ \bigr\} ~~\eqno[1]$$
where $\Phivec$ is an N-component
real scalar field, and we
denote $g_0\equiv \lambda_0 N$.

\noindent
-----------------------------------------------------------------------------
 {\tenrm\baselineskip=11pt

 \noindent *~Summary of a talk at the $3^{rd}$
 Workshop on Thermal Field Theories - Banff,
Canada ~-~Aug. 1993

\noindent ~{ Supported
in part by the  Bi-National Science Foundation (BSF),
the Henri Gutwirth Fund and the Fund for Promotion of
Research at the Technion }

\noindent $\dagger$~{ e-mail address : }
phr74mm@technion.bitnet}
\eject
\twelverm
\baselineskip=14pt
In the conventional treatment
of the large N limit  Eq. [1] is expressed by:
$$ Z_N  =\int {\cal D}\Phivec\int {\cal D}\sigma\int
{\cal D}m^2 exp\bigl\{-\int d^d x \,
\{ {1\over2}{(\partial_{\mu}
\Phivec)}^2 + {\mu^2_0\over 2} \sigma +
{\lambda_0 \over 4}\sigma^2   -{m^2 \over 2}
(\sigma-\Phivec^2)\}~\bigr\} ~. \eqno[2]$$
The double scaling limit, $N \to
\infty$ and $g_0 \to g_{0c}$,
of the O(N) vector models results in an expansion
in inverse powers of N.
$$\ln Z_N~=~\sum_{G,b} N^{(1-G)}~~
({N\over\beta})^b ~~F_b
\eqno[3]$$
There is a clear classification of the
Feynman graphs contributing to
each power  of $N^{-G}$ and the
expansion  resembles  an expansion  of randomly branched
polymers$^{\ambj-\zinn}$.
Following  here the suitable scaling procedure,
Eq. [3] turns into
an expansion in  a "polymer" coupling  constant,
where its  criticality is due to the potential and the
centrifugal barrier in Eq.[1].

Eq.[3] is the analog of the genus expansion
in matrix models.
The  double scaling limit in large N matrix models in
zero dimensions provides
a nonperturbative treatment of  string theory $^{\brez}$.
This  limit
is taken in the calculation of the partition function
$$ Z_N(g)~=~\int {\cal D}\hat\Phi
e^{-\beta Tr\{U(\hat\Phi)\} }  \eqno [4]$$
where $\hat\Phi$ is an NxN Hermitian matrix ,
U($\hat\Phi$) is the potential,
depending on the coupling constant(s) g.
The measure ${\cal D}
\hat\Phi$ can be  written in the form
$$ {\cal D}\hat\Phi~=~\prod_{i}^{N}d\phi_{ii}
\prod_{i<j}dRe\phi_{ij} dIm\phi_{ij} 
{}~=~\prod_{i}^{N}d\lambda_i \prod_{i<j}
(\lambda_i- \lambda_j
)^2 d\omega \eqno [5]$$
After performing the integration on
the U(N) angle variables $d\omega$,
one is left with the  integration on
the eigenvalues $\lambda_i$
$$ Z_N(g)~=~\Omega_N\int\prod_{i}^{N}d\lambda_i
{}~~exp \bigl\{2\sum_{i,j} \ln |\lambda_i- \lambda_j | - \beta
\sum_{i}U(\lambda_i)\bigr\}. \eqno [6]$$
In Eq. [6] one notes a Pauli repulsion
between the eigenvalues, and  a
critical point $g=g_c$ is found, when
the Fermi level reaches the extremum of
the potential $^{\brez}$. The genus expansion
of the fixed area partition function is:
$$\ln Z_N~=~a~+~b\ln\beta~+~\sum_{G,S} N^{2(1-G)}~
{}~({N\over\beta})^A ~~F_S
\eqno[7]$$
 Following the suitable scaling procedure in
 the double scaling limit; N=$\beta
\rightarrow \infty$ and $g \rightarrow g_c$ in a
correlated way, Eq. [7] turns into
an expansion in the string coupling  constant since every
genus is relevant now.

Returning now to the O(N) vector model, the singularity
structure in the coupling constant $g_0$ of each term
in the expansion is determined from the leading
$1 \over N $ term. In order to calculate the leading term in the
effective action one
 integrates  out first $\tilde{\vec\phi}(x)~,~$(where~$\vec\Phi(x)=
\vec\phi_c+\tilde{\vec\phi}(x)~)$
and  $\sigma(x)$. The $m^2(x)$ integration
is carried out by a saddle point integration. The effective
action $S_{eff}\{\phi_c\}$
is then obtained in the large N limit. $L^{-d}S_{eff}\{\phi_c\}$
 is proportional to the free  energy per
unit volume${^{\brezA}}$  at fixed
$ \vec\phi_c = L^{-d}\int d^dx \Phivec(x) .$

One finds:
$$ \eqalign{ e^{ -S_{eff}\{\phi_c\} }
&=C e^{-{N L^d\over 2}\bigl\{ \int\ddp
\ln ~( p^2 + m^2 )
 ~- {( 2 N \lambda_0 )}^{-1} (m^2-\mu_0^2)^2
{}~+ m^2\bigl( {\vec\phi_c^2\over N}\bigr)\bigr\}  }  \cr  &
\int {\cal D}\alpha(x)
e^{ -{N\over 2} \bigl\{  \int d^dx \alpha(x)
\bigl( \int\ddp
{1\over {p^2+m^2}} ~-~ {(  N \lambda_0 )}^{-1} (m^2-\mu_0^2)
{}~+~\bigl ( {\vec\phi_c^2\over N}\bigr)  ~\bigr)
\bigr\}  } \cr & \hskip 1.5 cm  e^{ {N \over 4}
 \bigl\{ \int dx dy \alpha(x) \alpha(y)
\int \ddp e^{ip(x-y)} \bigl( \Sigma(p)
+ {1 \over {N\lambda_0}} \bigr)
{}~~~+~~~ O(\alpha^3) \bigr\}  }   }   \eqno[8]$$
where $m^2(x)=m^2+\alpha(x)$ and $m^2=m^2(\phivec^2_c)$
is the solution of the gap
equation-saddle point condition:
$$ m^2 = \mu^2_0 + \lambda_0 N\biggl\{\int
{d^dk \over (2\pi)^d}
\bigl\{{1\over k^2 + m^2}\bigl\} ~~+~~
{\vec\phi_c}^2 \biggr  \} \eqno[9]$$
where  ${\vec\phi_c}^2$ has been rescaled by a factor of N.

The double scaling limit$^{\ambj-\zinn}$ is reached in
the limit at which  $N \to \infty$
is correlated with the limit $\lambda_0 \to \lambda_{0c}$,
where $\lambda_{0c}$
is the value of the coupling constant at which the
$O(\alpha^2)$ term at p=0 in Eq.[8] vanishes
(This can be formulated either
in terms of the unrenormalized theory
or in the renormalized theory, where
$\lambda_{Renorm.}\equiv\lambda \to \lambda_{c}$) . Namely,
$$\Sigma(p=0) + {1 \over {N\lambda_{0c}}} ~=~0~\eqno[10]$$
where
$$ \Sigma(p) = \int \ddk~
{1\over [(k-p)^2 + m^2]~[k^2 + m^2]}$$

In physical terms, the double scaling limit is reached when
the self coupling $\lambda_0$ of $\Phivec$
reaches the value $\lambda_{0c}$, at which the strength
of the force between the quanta of
the $\Phivec$ fields, the fundamental
scalars  in the theory, binds
them together strongly enough to create
a {\bf massless} scalar, O(N)singlet,
$\Phivec\cdot\Phivec$ bound state.
One also notes that the ground state is O(N)
symmetric and  the physical mass $m^2$ of the $\Phivec$
scalars is the solution of
the gap equation at $\phivec^2_c=0$,
namely, $m^2=m^2(\phivec^2_c=0)$
(The minimum of the effective
action is at $\phivec^2_c=0$).

Following the above procedure, it has
been shown in Ref.$\schn$ that in
d=4 the effective potential is complex  and
the critical vector model
cannot be consistently defined. This property is closely
related to the triviality of $\lambda \Phivec^4$ in d=4 in the
large N limit$^{\bm}$. Namely,  at large N, the positively
coupled  theory $( \lambda_0 > 0 )$ is trivial,
while the negatively coupled
theory $( \lambda_0 < 0 )$ is inconsistent
for  large enough $|\lambda_0|$.
It has been also  shown$^{\divec}$
that  in d=2 a similar problem occurs,  associated
with the absence of massless bound states in two dimensions.

In fact, it can be shown  that  for any d
there is no consistent theory at
$ \lambda_0 \leq  \lambda_{0c} < 0$.
For this purpose it is enough to
consider the unrenormalized theory where the
gap equation and the criticality conditions are:
$$ m^2 = \mu^2_0 +g_0 [~I(m^2)~+~~
{\vec\phi_c}^2 ] ~~~{\rm and}~~~
  \Sigma(p=0) = {1 \over g_0}  \eqno[11]$$
where $I(m^2)$  denotes (at  finite ultraviolet cutoff):
$$I(m^2)=\int
{d^dk \over (2\pi)^d} \bigl\{
{1\over k^2 + m^2}\bigl\}~~~. \eqno[12]$$
We have
${1 \over g_0}=
{\del I(m^2) \over \del m^2}\bigl|_{m^2=m^2(0)}$
and $\mu_0^2=
m^2(0)-\bigl[I(m^2)/{\del I(m^2) \over \del m^2}\bigr]
\bigl|_{m^2=m^2(0)}$,
where $m^2(0)$ is the
solution of the gap equation at $\phi^2_c=0$.

Inserting $g_0$ and $\mu_0$ into the gap equation one finds:
$$[m^2(\phi^2_c)-m^2(0)]\bigl[  1+ \bigl({I(m^2(\phi^2_c))-
I(m^2(0)) \over m^2(\phi^2_c)-m^2(0)}\bigr){1\over {\del I(m^2)
\over \del m^2}\bigl|_{m^2=m^2(0)} }\bigr] =
{\phi^2_c \over {\del I(m^2)
\over \del m^2}\bigl|_{m^2=m^2(0)}}
 \eqno[13]$$

One notes however that for any d and
fixed ultraviolet cutoff we have:
$$ I(m^2)>0~~,~~{\del I(m^2) \over \del m^2} < 0~~,~~
{\del^2 I(m^2) \over \del^2 m^2}
< 0 ~~~etc.\eqno[14]$$
and thus, there is no real solution
for $m^2(\phi^2_c)$ in Eq[13].
A similar result is obtained for the renormalized theory at
$\lambda=\lambda_c$.

The emerging physical picture  for
the double scaling limit
is  clear. In this limit, the force between the fundamental
quanta is set to the
value at which the  mass  of their bound state
vanishes.  This requires, however, tuning the coupling
constant to a negative value at which the theory is inconsistent.
Vacuum fluctuations
avoid giving a physical meaning to the large N theory
at the value of the coupling constant
$\lambda_0=\lambda_{0c}<0$. In the scalar theory the instability of
this limit is manifested by the absence of a real solution
to the relevant equations,
namely, the absence of a real effective action.
In the case of asymptotically free theories (in d=2
and probably at d=4 as well$^{\divec}$) the double scaling critical
condition is not satisfied for any value of the coupling constant.
In any case, a formal analytical
continuation is required in order to give a
physical meaning to the double scaling limit$^\mosh$.

A possible definition of the double scaling limit can be accepted
starting with
$\lambda_0 > \lambda_{0c}$  and  large N. In this region
of parameters, a
metastable ("false vacuum") exists whose lifetime$^{\bm,\thooft}$
is of the order of
$exp[O(N)]$. A suitable limit can be now taken at which
the theory is well
defined at any large but finite N and at any finite
$\epsilon \equiv \lambda_0 -
\lambda_{0c}$. The double scaling limit is now
taken while $N \to \infty$
and $\lambda_0 \to \lambda_{0c}$ in a
correlated manner (namely, taking
the cutoff to infity and  $N \to \infty$
in a correlated manner$^\neub$).

At finite temperature the
above procedure requires extra attention
due to the presence of the negative coupling constant.
$I(m^2)$ in Eq.[12] is replaced by:
$$I(m^2,T)=\int
{d^dk \over (2\pi)^d} \bigl\{{1\over k^2 + m^2} +
{2\pi\delta(k^2+m^2)
\over e^{E_k/T}-1}\bigl\}~~~. \eqno[15]$$
One finds now$^\eyal$ that even if  we start with
$0 > \lambda_0  >
\lambda_{0c}$ at some given temperature, namely, in
a metastable vacuum as in the
T=0 case, the metastable vacuum
disappears as the temperature
is increased$^\bm$ beyond some $T=T_c$.
The decay of the false vacuum
as the temperature is raised can be traced
to the fact that $\lambda_0  < 0$.
\vglue 0.6cm
\centerline{\twelvebf Acknowledgements}
\vglue 0.4cm

I would like to  thank the organizers of the workshop
at Banff and the theory group
at FERMILAB for their hospitality
during  last summer. I also  thank A. Duncan and
J. Feinberg for  useful discussions.

\vfill
\vskip 2 true cm
\bigskip
\centerline {\bf References}
\vskip 1 true cm

\def\PHL#1#2#3{{\it Phys. Lett.} {\bf#1B} #2 (19#3) }
\def\PRL#1#2#3{{\it Phys. Rev. Lett.} {\bf #1} #2 (19#3) }

\def\PRD#1#2#3{{\it Phys. Rev. }{\bf D#1} #2 (19#3)}
\def\NUP#1#2#3{{\it Nucl. Phys.} {\bf B#1} #2 (19#3)}

\itemitem{1.} J. Ambjorn,
B.Durhuus and T. J\'onsson , \PHL {244}{403}{90}
\itemitem{2.} S. Nishigaki and T. Yoneya , ~\NUP {348}{787}{91}
\itemitem{} A.Anderson,  R.C. Myers and V. Periwal, \PHL{254}{89}{91}
{}~\NUP {360}{463}{91}.
\itemitem{} P. Di Vecchia, M. Kato
and N. Ohta,  ~\NUP  {357}{495}{91}.
\itemitem{3.} J. Zinn-Justin, \PHL {257}{335}{91}
\itemitem{} P. Di Vecchia,
M. Kato and N. Ohta,  {\it~Int.  Journal of Mod.
Phys.} {\bf 7A} 1391 (1992).
\itemitem{4.} E. Br\'ezin and V.A. Kazakov , \PHL {236} {144} {90}
\itemitem{}M.R. Douglas and S.H. Shenker,  \NUP {335} {635} {90}
\itemitem{}D.J. Gross and A.A. Migdal,
\PRL {64}{127}{90} ;   \NUP {340}{333}{90}
\itemitem{5.}H. Neuberger, ~\NUP {340}{703}{90}
\itemitem{6.}See e.g. E. Br\'ezin and
J. Zinn-Justin ~\NUP{257}{867}{85}
\itemitem{7.} H. J. Schnitzer,
{\it Mod. Phys. Lett. }{\bf A7 } 2449 , (1992).
\itemitem{8.}W.A. Bardeen and M. Moshe, ~\PRD {28}{1372}{83}\lb
{}~and~ \PRD {34}{1229}{86}.
\itemitem{9.} P. Di Vecchia and M. Moshe \PHL {300}{49}{93}
\itemitem{10.} M. Moshe - Proceedings of
Renormalization Group '91 Conference,
Dubna, Sept. 1991, Ed. D. V.
Shirkov, World Scientific Pub. 1992.
 \itemitem{11.} G. t' Hooft ~
\PHL {109}{474}{82} ; {\it Commun.
Math. Phys.} {\bf 86}, 449 (1982)
\itemitem{12.} G. Eyal ~-~ M.Sc. thesis -
Technion - to be published.

\end